\documentclass[aps,prl,twocolumn,superscriptaddress,floatfix]{revtex4}
\usepackage[dvipdfmx]{graphicx}
\usepackage[dvipdfmx]{color}
\usepackage{amsmath}
\usepackage{amssymb}
\usepackage{braket}
\usepackage{dcolumn}
\usepackage{bm}
\usepackage{latexsym} 
\usepackage{setspace}
\usepackage{graphicx}
\usepackage{color}
\begin{document}
\newcommand{\h}{\mathcal{H}}
\newenvironment{e}{\begin{equation}}{\end{equation}}
\title{Spin-current coherence peak in superconductor/magnet junctions}
\author{Maki Umeda}
\email{maki.sssk@imr.tohoku.ac.jp}
\affiliation{Institute for Materials Research, Tohoku University, Sendai 980-8577, Japan}
\author{Yuki Shiomi}
\altaffiliation[Present address: ]{Department of Applied Physics and Quantum-Phase Electronics Center (QPEC), University of Tokyo, Hongo, Tokyo 113-8656, Japan, and RIKEN Center for Emergent Matter Science (CEMS), Wako 351-0198, Japan}
\affiliation{Institute for Materials Research, Tohoku University, Sendai 980-8577, Japan}
\author{Takashi Kikkawa}
\affiliation{Institute for Materials Research, Tohoku University, Sendai 980-8577, Japan}
\affiliation{Advanced Institute for Materials Research, Tohoku University, Sendai 980-8577, Japan}
\author{Tomohiko Niizeki}
\affiliation{Advanced Institute for Materials Research, Tohoku University, Sendai 980-8577, Japan}
\author{Jana Lustikova}
\affiliation{Institute for Materials Research, Tohoku University, Sendai 980-8577, Japan}
\author{Saburo Takahashi}
\affiliation{Institute for Materials Research, Tohoku University, Sendai 980-8577, Japan}
\affiliation{Center for Spintronics Research Network, Tohoku University, Sendai 980-8577, Japan}
\author{Eiji Saitoh}
\affiliation{Institute for Materials Research, Tohoku University, Sendai 980-8577, Japan}
\affiliation{Advanced Institute for Materials Research, Tohoku University, Sendai 980-8577, Japan}
\affiliation{Center for Spintronics Research Network, Tohoku University, Sendai 980-8577, Japan}
\affiliation{Advanced Science Research Center, Japan Atomic Energy Agency, Tokai 319-1195, Japan}
\date{\today}
\begin{abstract}
Coherence peak effects in a superconductor induced by a thermal spin current are reported. We measured inverse spin Hall effects induced by spin injection from a ferrimagnetic insulator Y$_3$Fe$_5$O$_{12}$ into a superconductor NbN using longitudinal spin Seebeck effects.  In the vicinity of the superconducting transition temperature of the NbN, a large enhancement of the spin Seebeck voltage is observed, whose sign is opposite to that for the vortex Nernst effect, but is consistent with a calculation for a coherence peak effect in the superconductor NbN.  
\end{abstract}
%
%\pacs{75.70.-i, 75.47.-m, 85.75.-d}% PACS, the Physics and Astronomy
%
%75.70.-i: Magnetic properties of thin films, surfaces, and interfaces
%72.25.-b: Spin polarized transport
%75.47.-m: Magnetotransport phenomena; materials for magnetotransport
%85.75.-d: Magnetoelectronics; spintronics: devices exploiting spin polarized transport or integrated magnetic fields 
%
\maketitle
%
%------------main-text----------------------------------
%
%%%%%%%%%%%%%%%%%%%%%%%%%%%%%%%%%%%%%%%%%%%%%%%%%%%%
%\section{I.~~INTRODUCTION}
%%%%%%%%%%%%%%%%%%%%%%%%%%%%%%%%%%%%%%%%%%%%%%%%%%%%
%
 Superconducting spintronics is an emerging research field which explores new spintronic functions by combining superconducting and magnetic orders \cite{SCspintronics1,SCspintronics2,SCspintronics3,SCspintronics4,SCspintronics5}. One of the key ingredients in the superconducting spintronics is generation of spin-polarized carriers in superconductors \cite{spinpolarizedtransport}. In conventional {\it s}-wave superconductors, a spin-singlet condensate does not carry spin angular momentum. However, thermally excited quasiparticles (QPs) can carry spin angular momenta even in {\it s}-wave superconductors \cite{SCspintransport}. It was theoretically proposed that spin currents due to the spin-polarized QPs are deflected by spin-orbit scattering to yield a charge imbalance along the Hall direction \cite{SCspinHall}, which can be detected using the technique of the inverse spin Hall effect (ISHE) \cite{spinHalleffect,inversedspinHall,spininjectionNbN}. A giant ISHE due to the QP spin current was indeed observed recently by electrical spin injection into an {\it s}-wave superconductor NbN in a  lateral spin valve structure \cite{quasiparticlespinHall}.
\par

A spin Seebeck effect (SSE) is one of the useful ways to generate spin current in magnetic heterostructures by applying a temperature gradient \cite{spinSeebeck}. In bilayer systems made up of a ferro- or ferri-magnet (F) and a nonmagnetic metal (N), the SSE enables spin injection from the F layer into the attached N layer through a thermal spin pumping originating in a non-equilibrium spin dynamics \cite{thermalspinpumping}. Several theories have been proposed for the mechanism of the SSE \cite{thermalspinpumping, linearresponseSSE, SSEtheory, SSE1}. Adachi {\it et al.} formulated the SSE using the linear-response theory \cite{linearresponseSSE, SSEtheory}, where the spin current flowing across the N/F interface reflects spin susceptibilities of both layers and the interface $s$-$d$ exchange coupling. The injected spin current into the N layer is converted into charge current by ISHE; the ISHE voltage arises in the direction of $\bm{j}_{s}\times\bm{\sigma}$, where $\bm{j}_{s}$ is the spin current and $\bm{\sigma}$ is the spin-polarization vector of electrons in the N layer. Hence ISHE voltage may reflect the spin dynamics of spin-detection layers in SSE measurements according to the theory \cite{linearresponseSSE}.
\par 
In this study, we have investigated ISHE induced by SSE in a superconductor/ferrimagnet (S/F) bilayer comprising a superconductor NbN and a ferrimagnetic insulator Y$_3$Fe$_5$O$_{12}$ (YIG). 
%In the normal state of the NbN layer, the ISHE voltage induced by the SSE is clearly observed, while it disappears in the zero-resistivity state. 
We show that the generated ISHE voltage exhibits an anomalous enhancement just below the superconducting transition. The voltage enhancement can be attributed to a coherence peak effect \cite{textsuperconductivity}; according to a theoretical calculation, singularity in the QP density of states at superconducting transition temperature ($T_c$) leads to an enhancement of spin dynamics \cite{spinpumpingSC} in full-gap superconductors. 
%It was first established by measurements of nuclear magnetic resonance \cite{Hebelslichter}.
The coherence peak effects have been studied by measuring nuclear magnetic resonance \cite{Hebelslichter} and ac conductivity \cite{skineffect,conductivity}, but the present results show that ISHE is a useful tool for detection of coherence effects in superconductors.

%In {\it s}-wave superconductors, it has been known that an enhancement (coherence peak) in the dynamic spin susceptibility appears just below $T_c$ in the scattering and absorption of external field \cite{textsuperconductivity}, as observed in the NMR relaxation rate measurement \cite{Hebelslichter}.
%Since the sign of the enhanced voltage is the same as that of ISHE voltage, but opposite to that of electromotive-force induced by vortex flow (vortex Nernst effect), the enhanced voltage signal is attributed to ISHE induced by the SSE. 
%Our theoretical calculation based on the linear response theory shows that the ISHE in the S/F bilayer system is enhanced just below $T_c$ owing to a coherence peak effect, consistent with the experimental observation. 
\par

%
%%%%%%%%%%%%%%%%%%%%%%%%%%%%%%%%%%%%%%%%%%%%%%%%%%%%
%\section{II.~~METHODS} \label{sec:procedure}
%%%%%%%%%%%%%%%%%%%%%%%%%%%%%%%%%%%%%%%%%%%%%%%%%%%%

\begin{figure}[htb]
\begin{center}
\includegraphics{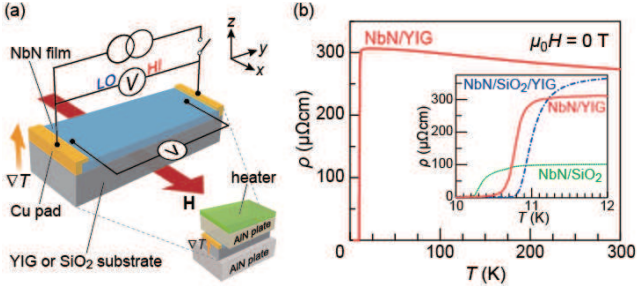}
\caption{(a) A schematic illustration of the longitudinal SSE measurement. Ohmic contacts are made at the Cu pads. The temperature gradient $\nabla T$ and magnetic field $H$ are applied in the $+z$ and $\pm x$ directions, respectively, and the thermoelectric voltage is measured in the $+y$ direction. Resistivity ($\rho$) measurement is done with a four-point probe. (b) Temperature ($T$) dependence of $\rho$ for NbN/YIG (red color), NbN/SiO$_2$(100 nm)/YIG (blue color), and NbN/SiO$_2$ (green color)  in zero magnetic field. }
\label{1}
\end{center}
\end{figure}

 As a F layer, we chose a (111) YIG slab available commercially. A band gap of YIG is 2.85\ eV much larger than $\sim$\ 4\ meV of superconducting gap of NbN \cite{YIGenergygap1,YIGenergygap2,NbNenergygap}. Thanks to this energy difference, a superconducting proximity effect is negligible in our measurements \cite{proximitySP}. The size of the YIG slab is as follows: the lengths along the $x$, $y$, and $z$ directions are $2.0$\ mm, $6.0$\ mm, and $1.0$\ mm, respectively  [see the $xyz$ coordinate in Fig. \ref{1}(a)]. The surface ($xy$-plane) of the YIG slab was mechanically polished, and then a 20-nm-thick NbN film was sputtered on it; here, the thickness is greater than the spin diffusion length of 7\ nm reported for a NbN film \cite{quasiparticlespinHall}. For the sputtering, a pure Nb (5N) target was used in a mixture of argon and nitrogen gases at room temperature  \cite{NbNreactivesputtering1,NbNreactivesputtering2}. The base pressure was better than $2.0\times10^{-5}$\ Pa.\par

The measurement of the SSE was performed using the longitudinal setup with a superconducting magnet \cite{criticalsuppression}, where an external magnetic field ($H$) up to $9$ T was applied along the $x$ direction in the film plane [$xy$-plane in Fig. \ref{1}(a)]. The sample was sandwiched by two aluminum-nitride plates. The temperature difference $\Delta T$ along the $z$ axis was applied using a chip resistance heater [Fig. \ref{1}(a)] and measured with a couple of type-E thermocouples. The ISHE voltage which arises in the $y$-direction due to $\nabla T$($||z$) was measured using a nanovoltmeter at each $H$ value. As shown in Fig. \ref{1}(a), a 30-nm-thick Cu film was deposited at ends of the NbN film, and ohmic contacts were made on the Cu parts for the measurement of ISHE voltage. To compare the ISHE voltage and the transport properties of the NbN layer under a temperature gradient, 4-wire resistance was measured under the same $\Delta T$ values. The resistance of the NbN layer was used also as a thermometer to determine the average sample temperature of the NbN layer under $\Delta T$. %$I$-$V$ curves were also measured with another nanovoltmeter and a source meter to determine the the $T$-$B$ phase diagram under $\Delta T$. 
\par  
%
%
%
%
%%%%%%%%%%%%%%%%%%%%%%%%%%%%%%%%%%%%%%%%%%%%%%%%%%%%
%\section{III.~~RESULTS}
%%%%%%%%%%%%%%%%%%%%%%%%%%%%%%%%%%%%%%%%%%%%%%%%%%%%
Resistivity, $\rho$, for the NbN/YIG bilayer measured without applying $H$ nor $\Delta T$ is shown as a function of system temperature, $T$, in Fig. \ref{1}(b). The magnitude of $\rho$ is about 270 ${\mu\rm \Omega cm}$ at $300$ K. $\rho$ increases slightly as $T$ decreases from 300 K, and suddenly drops around 11 K. %The occurrence of the zero resistivity state below 10.5 K clearly demonstrates that the NbN layer shows superconductivity below that temperature.
The resistivity becomes zero below 10.5 K, signifying superconductivity of the NbN layer.
$T_{c}$ is determined as the temperature where $\rho$ becomes 95 \% of the normal state resistivity  $\rho_n$. In the inset to Fig. \ref{1}(b), the $T$ dependences of $\rho$ for a NbN/SiO$_2$/YIG trilayer and a NbN/SiO$_2$ bilayer are also shown, where the thickness of the SiO$_2$ layer is $100$ nm. The NbN films in NbN/SiO$_2$/YIG and  NbN/SiO$_2$ show similar $T_c$: 11.7 K and 10.6 K, respectively. 
\par
\begin{figure}[t]
\centerline{\includegraphics{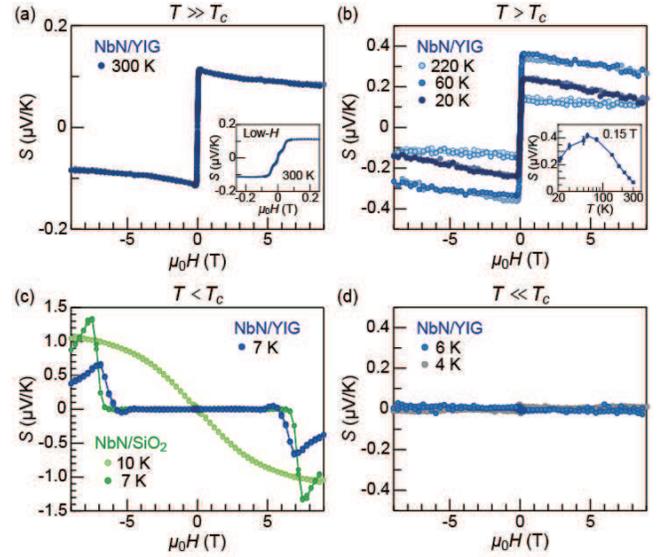}}
\caption{$H$ dependence of $S$ for NbN/YIG (a) at $T=300$\ K, (b) at selected temperatures above $T_c$, (c) at $7$ K just below $T_c$ (blue circles), and (d) in the zero-resistivity state below $T_c$. The $H$ was swept between $-9$ T and $9$ T. The inset to (a) shows a magnified view of $S(H)$ in the low-$H$ range
($|\mu_0H|\lesssim0.25\ \textrm{T}$) at $300~\textrm{K}$. The inset to (b) presents the $T$ dependence of the saturated value of $S$ at $\mu_0H=0.15$\ T. In (c), $H$ dependence of $S$ for NbN/SiO$_2$ at 7\ K and 10\ K is also shown (green and light green circles). }
\label{2}
\end{figure}
In Fig. \ref{2}, we show the experimental results of the SSE measurements, by plotting $S\equiv V/\Delta T$. Figure \ref{2}(a) shows the $H$ dependence of $S$ measured at 300 K, which is much higher than $T_c=11.1$ K. The observed $H$ dependence of $S$ is antisymmetric with respect to the magnetic field. By increasing $H$ from zero, $S$ gradually increases reflecting the magnetization in YIG [inset to  Fig. \ref{2}(a)], but is clearly suppressed by futher increasing $H$ ($\mu_0H>0.1$ T). This suppression of $S$ is consistent with the typical SSE feature \cite{criticalsuppression}; magnon excitations responsible for  SSE are suppressed by strong magnetic fields. The sign of $S$ is consistent with the spin Hall angle of Nb \cite{NbSHC}. As $T$ decreases in the normal-state regime far above $T_{c}$, the magnitude of $S$ shows a maximum around $60$ K and then decreases with decreasing $T$, as shown in Fig. \ref{2}(b).
\par
%
%As $T$ decreases in the normal-state regime far above $T_{c}$, the magnitude of $S$ increases and shows a maximum around $60$ K, as shown in Fig. \ref{2}(b). Below $60$ K in the low $H$ regime ($|\mu_0H|\leq0.2$ T), the magnitude of $S$ decreases with decreasing $T$ [inset to Fig. \ref{2}(b)]. The similar $T$ dependence of $S$ was reported for Pt/YIG \cite{criticalsuppression}; the SSE signal shows a maximum at around $70$ K. As $T$ decreases, the high-$H$ suppression of $S$ becomes more prominent, reflecting the fact that the field-induced freeze-out effect of
%magnons becomes more significant at lower temperatures \cite{LSSEYIGmagnondispersion}. 
\par
Figure \ref{2}(c) shows the $H$ dependence of $S$ for the NbN/YIG at 7 K, which is immediately below $T_c$. The magnitude of $S$ is almost zero in the low $H$ regime, but suddenly increases at around $\pm 6$ T. The onset $H$ corresponds to a superconducting critical field. A similar behavior is observed in the NbN/SiO$_2$ as shown in Fig. \ref{2}(c). In the normal state in the vicinity of the transition, large $\Delta T$-induced voltage signals are observed [see Figure \ref{2}(c)], which are due to a superconducting vortex flow by thermal gradients. This effect is known as the vortex Nernst effect (VNE), where the vortex flow in the direction of decreasing temperature gives rise to a voltage along the Hall direction \cite{vortexNernst1,vortexNernst2,vortexNernst3,vortexNernst4,vortexNernst5}. When the $H$ direction is reversed, the direction of the magnetic flux of vortices is reversed, and the sign of the voltage due to the VNE is reversed, consistent with the experimental results in Fig. \ref{2}(c).
Note that the sign of the VNE voltage is opposite to that of the spin Seebeck voltage, as shown in Figs. \ref{2}(a), (b), and (c). 
%; electric field induced by vortex Nernst effect is written as $\bm{E}=\mu_0\bm{H}\times\bm{v}_{\phi}$ where $\bm{v}_{\phi}$ is vortex flow velocity which is along $-z$ direction [Fig. \ref{1}(a)]. 
%With further increasing $H$ in the normal state, the superconducting fluctuation becomes smaller, and the VNE is suppressed.  
\par
As $T$ further decreases down to 4 K, which is much lower than $T_c$, no voltage signal is observed even up to $9$ T. At very low temperatures below $T_c$, the zero-resistivity state persists up to $9$ T, which means that vortices are strongly pinned in the NbN layer. Therefore,  the voltage due to the VNE disappears in a mixed state at low temperatures \cite{SCinTgradient}. The voltage due to SSE also disappears owing to the shunting effect due to the zero resistivity \cite{nonlocalspininjection}. The disappearance of the ISHE voltage in the zero-resistivity state of the NbN is in stark contrast to a previous report on the giant ISHE induced by electrical spin injection into a NbN layer \cite{quasiparticlespinHall}. In our case, the voltage contact is ohmic, different from the non-ohmic contact in the previous report \cite{quasiparticlespinHall}, and charge neutrality in the superconductor can be achieved by the cancellation of the QP charge currents and supercurrents \cite{textNonEquilibriumSuperconductivity}.
\par
\begin{figure}[t]
\centerline{\includegraphics{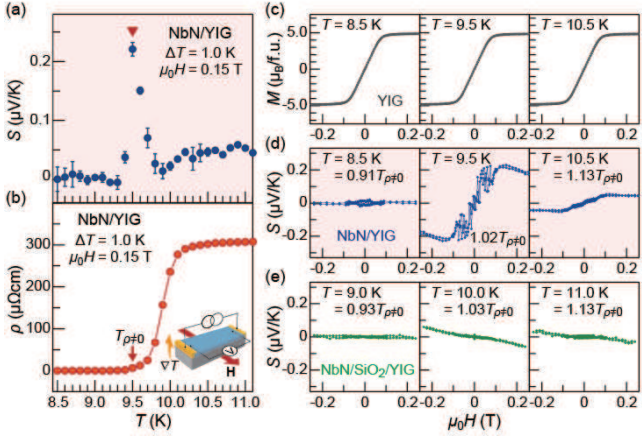}}
\caption{(a),(b) $T$ dependence of (a) $S$ and (b) $\rho$  for NbN/YIG near $T_c$ at $\mu_0H=0.15$\ T with $\Delta T=1$\ K. (c) $H$ dependence of magnetization $M$ for the YIG substrate at low temperatures. (d),(e) $H$ dependence of $S$ for (d) NbN/YIG and (e) NbN/SiO$_2$/YIG in the low-$H$ regime. Here, $H$ was swept between $-0.25 $ T and $0.25$ T. For comparison of results between the two samples, the data taken at similar $T/T_{\rho\neq 0}$ values are shown where $T_{\rho\neq 0}$ is the temperature at which $\rho$ begins to increase from the zero value.}
\label{3}
\end{figure}
To study SSE signals around $T_{c}$ in detail,  in Fig. \ref{3}(a), we show the $T$ dependence of $S$ around $T_c$ taken at $\mu_0H=0.15$ T where $\mu_0$ is the magnetic parmeability of vacuum. Here, $\Delta T$ is 1 K, and the sample average temperature shifts from the system temperature by about 1 K. The $H$ value is so small that $S$ is attributed simply to SSE. The positive sign of $S$ in Fig. \ref{3}(a) is consistent with the ISHE due to the SSE observed in the normal state above $T_{c}$ [see Figs. \ref{2}(a) and (b)]. 
Above 10.0 K, $S$ is first constant and then  decreases with decreasing $T$ at 10.2 K. On the other hand, below 9.3 K, $S$ is zero. In these two temperature regions, the $T$-dependence of $S$ is consistent with that of $\rho$ in NbN.
\par
%As $T$ decreases, the magnitude of $S$ is almost constant above $10.2$ K, but it starts decreasing at $10.2$ K, and finally becomes zero below $9.3$ K. This overall $T$ dependence of $S$ is consistent with that of $\rho$ measured for NbN [Fig. \ref{3}(b)]. 
\par
%
%Surprisingly, the magnitude of $S$ is strongly enhanced in a narrow $T$ range around $T_c$ [Fig. \ref{3}(a)], while the magnetization $M$ for YIG is almost unchanged in the entire temperature range, as shown in Fig. \ref{3}(c).
We found that, surprisingly, in a narrow temperature region 9.3 K$<T<$10.0 K, the magnitude of $S$ is strongly enhanced [Fig. \ref{3}(a)]. At the same time, the magnetization $M$ of YIG is almost unchanged in the entire temperature range as shown in Fig. \ref{3}(c).
The enhancement of $S$ immediately below $T_c$ cannot be explained by VNE, but is attributable to ISHE induced by SSE for the following reasons. First, the $H$ dependence of $S$ at 9.5 K almost traces the $M$ curve of YIG, as shown in Figs. \ref{3}(c) and (d), indicating that the $H$ dependence of $S$ is relevant to the YIG magnetization. Second, the sign of the enhanced $S$ signal is the same as that of the ISHE voltage induced by the SSE, but is opposite to that of the VNE voltage. Third, the enhancement of $S$ is not observed in NbN/SiO$_2$/YIG trilayer samples, where the SiO$_2$ layer blocks the transmission of spin currents from YIG to NbN, but VNE is still expected to appear. As shown in Fig. \ref{3}(e), on the other hand, negative voltage signals in positive magnetic fields are observed in NbN/SiO$_2$/YIG. $S$ is almost linear with respect to $H$ in the low-$H$ regime, consistent with the VNE observed in the NbN/SiO$_2$ bilayer at similar temperatures shown in Fig. \ref{2}(c).
\par
%%%%%%%%%%%%%%%%%%%%%%%%%%%%%%%%%%%%%%%%%%%%%%%%%%%%
%\section{IV.~~DISCUSSION}
%%%%%%%%%%%%%%%%%%%%%%%%%%%%%%%%%%%%%%%%%%%%%%%%%%%%
Very recently, spin pumping into superconductors has been theoretically studied \cite{spinpumpingSC}, and the temperature dependence of the spin pumping efficiency was predicted to exhibit a pronounced coherence peak immediately below $T_{c}$ \cite{YYao}. Following the theoretical approach of spin pumping into superconductors \cite{textspincurrent}, 
%we have performed a calculation of spin injection into a superconductor due to the SSE
we have calculated the spin current injected into a superconductor due to the SSE
in a superconductor/ferrimagnet (S/F) bilayer system. In the presence of $\Delta T$ between F and S layers, a spin current flows across the interface via SSE. Spin-flip scattering of QPs in the S layer takes place at the interface through the $s$-$d$ exchange interaction with localized moments in the F layer, which accompanies a magnon excitation in the F layer. Making use of the BCS-Bogoliubov theory \cite{textsuperconductivity} and linear response theory for the  fluctuation dissipation
theorem \cite{SSEtheory,linearresponseSSE}, we obtain the spin current $j^s_{\rm S/F}$ flowing across the S/F interface 
with magnon excitation with the frequency $\omega_q$,
\begin{widetext}
\begin{eqnarray}
j^s_{\rm S/F}(\omega_q, T) 
&\propto&
\left[n(\omega_q,T_{\rm F})-n(\omega_q,T_{\rm S})\right]
%\cr &
%\times
%&
\int_{-\infty}^\infty    
F({\it \Delta},E,\omega_{q})
N_S(E)N_S(E+\hbar\omega_q)
\left[f(E+\hbar\omega_q)-f(E)\right]dE,
\label{eq:Js-SF}
\end{eqnarray}
\end{widetext}
where $n(\omega,T)=1/(e^{\hbar\omega/k_{B}T}-1)$ is the Bose distribution function, $T_{\rm F}$ and $T_{\rm S}$ are the effective temperatures of magnons and QPs, respectively, ${\it \Delta}$ is the temperature-dependent superconducting energy gap, $E$ is the QP energy measured from the Fermi energy,  $N_S(E)=$Re$\left[|E|/\sqrt{E^2-{\it \Delta}^2}\right]$ is the normalized BCS density of states, and $f(E)=1/(e^{E/k_{B}T}+1)$ is the Fermi distribution function of QPs in the S layer.
The factor $
F({\it \Delta},E,\omega_q)=\frac{1}{2}\left(1+\frac{{\it \Delta}^2}{E(E+\hbar\omega_{q})}\right)\label{eq:coherence}$ represents a coherence effect of superconductivity for  spin-flip scattering of QPs \cite{textsuperconductivity}.
\par
\begin{figure}[t]
\centerline{\includegraphics{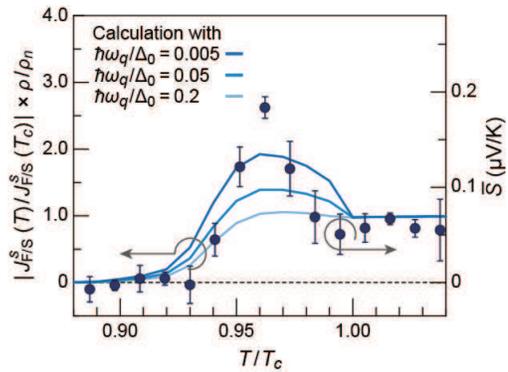}}
\caption{Comparison to a theoretical calculation. The results are shown as solid curves: $T/T_{c}$ dependence of $|j^{s}_{\rm S/F}(T)/j^{s}_{\rm S/F}(T_c)|$ multiplied by the measured resistivity $\rho/\rho_n$ (solid curves) is plotted for low magnon frequencies, $\hbar\omega_{\bm{q}}/{\it \Delta}_{0}=0.005, 0.05$, and $0.2$. $\omega_q$ and ${\it \Delta}_0$ are the magnon frequency and the magnitude of the superconducting energy gap at zero temperature, respectively. The experimental result is shown as filled circles: $T/T_{c}$ dependence of $\bar{S}$ (filled circles), which is defined by $[ V_{\mathrm{NbN/YIG}} - \delta s V_{\mathrm{NbN/SiO_2/YIG}}]/\Delta T$ with the adjusting parameter $\delta s=$0.92, is presented.}
\label{4}
\end{figure}

In Fig. \ref{4}, we show the numerical calculation of Eq.~(\ref{eq:Js-SF}) for several values of $\hbar\omega_q/{\it \Delta}_0$. For comparison of the calculation with the enhancement of $S$ observed experimentally [Fig. \ref{3}(a)], the normalized magnitude of the spin current $|j^{s}_{\rm S/F}(\omega_q, T)/j^{s}_{\rm S/F}(\omega_q, T_c)|$ is multiplied by $\rho/\rho_n$, and plotted as a function of $T/T_{c}$. Here, we used the values of $T_{c}=11$ K and ${\it \Delta}_0/k_\mathrm{B}=2.08T_c=23$ K where ${\it \Delta}_0$ is the superconducting gap at $T=0$ for the strongly coupled superconductor NbN \cite{textsuperconductivity2}. The temperature scale of $\it \Delta$ is much larger than that of the magnon gap energy of YIG estimated to be $g\mu_\mathrm{B}\mu_0H/k_\mathrm{B}=0.2$\ K at $\mu_0H = 0.15$\ T, where $g$ is the Lande's $g$ factor ($=2.0$ for YIG) and $\mu_{\mathrm{B}}$ is the Bohr magneton. Hence, we take into account the contribution only from low-frequency magnons in the calculation of Eq.~(\ref{eq:Js-SF}). As shown in Fig. \ref{4}, the calculated spin-current signal is strongly enhanced just below $T_{c}$ owing to the superconducting coherence effect, forming a coherence peak in the narrow $T$ range below $T_{c}$. For lower-frequency magnons ($\hbar\omega_{q}\ll{\it \Delta}$), a larger peak is expected to be produced by the coherence factor $F$ multiplied by the densities of states in Eq.~(\ref{eq:Js-SF}) \cite{textsuperconductivity,conductivityYBCO}.
\par
The experimental values for the SSE signal are calibrated as $\bar{S}\equiv \left[V_{\mathrm{NbN/YIG}}-\delta sV_{\mathrm{NbN/SiO_2/YIG}}\right]/\Delta T$, where the VNE signal multiplied by the adjusting parameter $\delta s$ is subtracted from $S$ \cite{adjustingparameter}. The result for $\delta s=0.92$ is plotted by filled circles  in Fig. \ref{4}. As seen in Fig. \ref{4}, the temperature range where the enhancement of $S$ observed experimentally is reproduced by the calculation; both experimental and theoretical results are largest at $T\approx$ 0.96 $T_c$. The coherence-peak temperature in the calculation is a bit higher than that reported in a microwave conductivity measurement for Nb ($0.8\sim0.9$ $T_c$) \cite{Nbcoherencepeak}. This may be because, in Fig. \ref{4}, the normalized spin current $|j^{s}_{\rm S/F}(T)/j^{s}_{\rm S/F}(T_c)|$ is multiplied by the normalized resistivity $\rho/\rho_n$ which rapidly decreases to zero below $T_{c}$. Since measured voltages are zero in the zero-resistivity state, the coherence-peak temperature shifts from $T\sim$ 0.9 $T_c$ expected in Eq.~(\ref{eq:Js-SF}) to $T\sim$ 0.96 $T_c$ in Fig. \ref{4}. The magnitude of  $\bar{S}$ enhancement is consistent with the theoretical calculation; compared to the normal-state values, the observed signal is enhanced by a factor of $\sim 2.5$ at $T/T_c \approx$ 0.96, and the enhancement in the calculation is $\sim2$ for $\hbar\omega_q/{\it \Delta}_0 = 0.005$. The small discrepancy in the magnitudes between the experimental and theoretical results could be due to impurity spin-orbit scattering, which may further enhance spin current owing to impurity vertex corrections \cite{spinpumpingSC}. 
\par
%

%%%%%%%%%%%%%%%%%%%%%%%%%%%%%%%%%%%%%%%%%%%%%%%%%%%%
%\section{V.~~CONCLUSION}
%%%%%%%%%%%%%%%%%%%%%%%%%%%%%%%%%%%%%%%%%%%%%%%%%%%%
%
%
In summary, we found that inversed spin Hall voltage induced by spin Seebeck effects in a bilayer film comprising NbN and YIG layers is clearly enhanced in a narrow temperature range immediately below $T_c$, exhibiting a peak structure. This enhancement appears only in NbN/YIG, not in NbN/SiO$_2$/YIG or NbN/SiO$_2$, where only vortex Nernst voltages whose sign is opposite to the inversed spin Hall voltages were observed. A theoretical calculation of the spin Seebeck effects reveals that a coherence peak effect in the spin Seebeck effects for superconductors in which low-frequency magnons are taken into consideration can be responsible for the anomalous enhancement of the spin Seebeck effects.      
\par

{\it Acknowledgment-} The authors thank K. Ohnishi and Y. Chen for fruitful discussions. This research was supported by JST ERATO ``Spin Quantum Rectification Project" (JPMJER1402), JSPS KAKENHI (No. 17H04806, No. JP16H00977, and No. 16K13827), and MEXT (Innovative Area ``Nano Spin Conversion Science" (No. 26103005)). T.K. and J.L. are supported by JSPS through a research fellowship for young scientists (No. JP15J08026 for T.K. and No. JP16J03699 for J.L.). M.U. is supported by GP-Spin at Tohoku University.
\par


\begin{thebibliography}{99}
\bibitem{adjustingparameter} The value of the adjusting parameter $\delta s$ is chosen so that
subtracting the pure VNE signal in NbN/SiO$_2$/YIG [Fig. \ref{3}(e)] from the
signal containing both SSE and VNE in NbN/YIG [Fig. \ref{3}(d)] removes the
negative VNE contribution at fields $\mu_0H=0.2$\ T.
\end{thebibliography}
\end{document}